\documentclass[aps,preprint,superscriptaddress,amsfonts,amssymb,amsmath]{revtex4}

\usepackage{epsfig}

\newcommand{\be}{\begin{eqnarray}}
\newcommand{\ee}{\end{eqnarray}}
\newcommand{\bm}{\boldsymbol}

\newcommand{\n}{\nonumber}

\begin{document}

\preprint{Yukawa Institute Kyoto}
\preprint{YITP-12-62}

~
\vskip 2 truecm

\title{Generalized Rayleigh and Jacobi processes \\
and exceptional orthogonal polynomials}

\author{C.-I. Chou}
\affiliation{Department of Physics, Chinese Culture University, Taipei 111, Taiwan. R.O.C. }

\author{C.-L. Ho}
\affiliation{Yukawa Institute for Theoretical Physics, Kyoto University, Kyoto 606-8502, Japan}
\affiliation{Department of Physics, Tamkang University,
 Tamsui 251, Taiwan, R.O.C.\footnote{Permanent address}}

\date{Jul 25, 2012}

\begin{abstract}

We present four types  of infinitely many exactly solvable Fokker-Planck equations, which are related to the newly discovered
exceptional orthogonal polynomials.  They represent the deformed versions of the Rayleigh process and the Jacobi process. 

\end{abstract}




 \maketitle




\section{Introduction}


The Fokker-Planck equation (FPE) is one of the basic equations
widely used for studying  the effect of fluctuations in
macroscopic systems \cite{RIS}. 
It has been employed in many
areas such as astrophysics
\cite{NM,HCDHQS}, chemistry \cite{NOW,LHE},
biology \cite{SG,FBF}, finance \cite{FPR} and
others. Because of its broad applicability, it is therefore of
great interest to obtain solutions of the FPE for various physical
situations.  Many methods, including analytical, approximate and
numerical ones, have been developed to solve the FPE 
\cite{RIS,BS,MP,BM,HU,LLPS,LAN}.  
More recently, the exact solvability of the FPE with both time-dependent drift and
diffusion coefficients has been studied by the similarity method \cite{LH2}.

As with any other equation, exactly solvable FPE's are hard to come by. 
Thus any new addition to the stock of exactly solvable FPE's is always welcome.
One of the methods of solving FPE with time-independent diffusion
and drift coefficients is to transform the FPE into a
time-independent Schr\"odinger equation, and then solve the
eigenvalue problem of the latter \cite{RIS,HS1,HD}. The
transformation to the Schr\"odinger equation of a FPE eliminates
the first order spatial derivative in the FPE and creates a
Hermitian spatial differential operator.  Owing to this connection, an exactly solvable
Schr\"odinger equation may lead to a corresponding exactly solvable FPE. In fact, the well-known solvable 
FPE describing the Ornsterin--Ulenbeck process  is related to the Schr\"odinger equation of the harmonic 
oscillator \cite{RIS}.   Another interesting process, called the Rayleigh process, is described by a FPE transformable
 to the Schr\"odinger equation of the radial oscillator \cite{GAR}.  
 
Based on this relation, we would like to present infinitely many new exactly solvable FPE's 
which are related to the quantal systems associated with the newly discovered polynomials, 
the so-called exceptional orthogonal polynomials (\cite{GKM1}-\cite{PTV}).  The discoveries of these new kinds of polynomials, and the
 quantal systems related to them, have been among the most interesting developments in mathematical physics
in recent  years.
 Unlike the classical
orthogonal polynomials, these new polynomials have the remarkable
properties that they  start with degree $\ell=1,2\ldots,$
polynomials instead of a constant, and yet they still 
 form complete sets with respect to some
positive-definite measure.

A brief remark on the main developments of these new polynomials is in order.  
For a recent review, see eg. \cite{Que4}.
Two families of such polynomials, namely, the Laguerre- and
Jacobi-type $X_1$ polynomials, corresponding to $\ell=1$, were
first proposed by G\'omez-Ullate et al. in Ref.~\cite{GKM1},
within the Sturm-Lioville theory, as solutions of second-order
eigenvalue equations with rational coefficients. The results in
Ref.~\cite{GKM1} were reformulated in the framework of quantum
mechanics and shape-invariant potentials by Quesne et al.
\cite{Que1,Que2}. These quantal systems turn out to be rationally
extended systems of the traditional ones which are related to the
classical orthogonal polynomials. The most general $X_\ell$
exceptional polynomials, valid for all integral $\ell=1,2,\ldots$,
were discovered by Odake and Sasaki \cite{OS1} (the case of
$\ell=2$ was also discussed in Ref.~\cite{Que2}). Later, in
Ref.~\cite{HOS} equivalent but much simpler looking forms of the
Laguerre- and Jacobi-type $X_{\ell}$ polynomials were presented.
Such forms facilitate an in-depth study of some important
properties of the $X_\ell$ polynomials, such as the actions of the
forward and backward shift operators on the $X_{\ell}$
polynomials, Gram-Schmidt orthonormalization for the algebraic
construction of the $X_{\ell}$ polynomials, Rodrigues formulas,
and the generating functions of these new polynomials. Structure
of the zeros of the exception polynomials was studied in
Ref.~\cite{HS2,GMM}.  Recently, these polynomials were
constructed by means of the Darboux-Crum transformation
\cite{Que2,GKM2,STZ}.  An approach which does not rely on 
Darboux-Crum and shape invariance was reported in \cite{Ho2}.
Generalizations
of these new orthogonal polynomials  to multi-indexed cases were
discussed in Ref.~\cite{GKM4,OS3}. Radial oscillator
systems related to the exceptional Laguerre polynomials have been
considered based on higher order supersymmetric quantum mechanics
\cite{Que3}. 

So far  most of these studies concern mainly with the generation and mathematical 
structure of the new polynomials.  Very few physical applications of these polynomials have been
studied. In this regard, we mention that the role of these new
polynomials was considered in \cite{MR} for the Schr\"dingier equation with a 
position-dependent mass, and in Ref.~\cite{Ho1} for the
Dirac equations coupled minimally and non-minimally with an electromagnetic field.  
Here we show that infinitely many exactly solvable FPE's can be constructed with these new polynomials.
They represent the generalized, or deformed versions of the Rayleigh and Jacobi processes.

The plan of the paper is as follows.   In Sect.~II we briefly review the transformation between FPE and the Schr\"odinger
equation, and derive  the general expression for the probability density function of the FPE which is related to the exceptional orthogonal polynomials. 
Sect.~III and IV then discuss the generalized Rayleigh process and the Jacobi process, respectively. 
Sect.~V concludes the paper.




\section{Fokker-Planck and Schr\"odinger equations}

In one dimension, the FPE of the probability density function (PDF) 
$\mathcal{P}(x,t)$ is \cite{RIS}
\begin{gather}
\frac{\partial}{\partial t} \mathcal{P}(x,t)=\left[-\frac{\partial}{\partial x} D^{(1)}(x) +
\frac{\partial^2}{\partial x^2}D^{(2)}(x)\right]\mathcal{P}(x,t). \label{FPE}
\end{gather}
The functions $D^{(1)}(x)$ and $D^{(2)}(x)$ in the FPE are, respectively,  the drift and the diffusion
coefficient (we consider only time-independent case).  The drift
coefficient represents the external force acting on the particle,
while the diffusion coefficient accounts for the effect of
fluctuation.   Without loss of generality, in what follows we
shall take $D^{(2)}(x)=1$.  The drift coefficient can be defined by a
drift potential $U(x)$ as $D^{(1)}(x)=-U^\prime (x)$, where the prime denotes derivative with respective to $x$. 
The current density $J(x,t)$ is given by
\begin{equation}
J(x,t)= -\frac{\partial U(x)}{\partial x}\, P(x,t) - \frac{\partial}{\partial x} P(x,t).
\label{J}
\end{equation}

The FPE is closely related to the Schr\"odinger equation
\cite{HD,HS1,RIS}.  To show this it is convenient to define a function $w(x)=-U(x)/2$, called the prepotential, so that 
$D^{(1)}(x)=2w^\prime(x)$.  It is the prepotential that occurs naturally in the Schr\"odinger equation.  
Substituting
 \be
 \mathcal{P}(x,t)\equiv e^{-\mathcal{E} t} e^{w(x)} \phi(x)
 \ee
into the FPE, we find that $\phi$ satisfies the
equation $H\phi=\mathcal{E}\phi$, where
\begin{eqnarray}
H
=-\frac{\partial^2}{\partial x^2}+w^\prime (x)^2 +
w^{\prime\prime}(x).\nonumber
\end{eqnarray}
Thus $\phi$ satisfies the time-independent Schr\"odinger equation
with Hamiltonian $H$ and eigenvalue $\mathcal{E}$, and
$\phi_0=\exp(w)$ is the zero mode of $H$: $H\phi_0=0$.

This implies that a FPE transformable to an exactly
solvable Schr\"odinger equations can be exactly solved. 
One needs only to link the prepotential $w(x)$ in the Schr\"odinger
system with the drift
coefficient $D^{(1)}(x)=2w^\prime (x)$ in the FPE.  For example, the
shifted oscillator potential in quantum mechanics corresponds to
the FP equation for the well-known Ornstein--Uhlenbeck process
\cite{RIS}.  

Specifically, suppose all the
eigenfunctions $\phi_n$ ($n=0,1,2,\ldots$) of $H$ with eigenvalues
$\mathcal{E}_n$ are solved, then $\mathcal{P}_n(x,t)=e^{-\mathcal{E}_n t}\phi_0(x)\phi_n(x)$ ($n=0,1,2,\ldots$)
 are solutions of the FPE (\ref{FPE}).
The stationary distribution, corresponding to $\mathcal{E}_0=0$,  is $\mathcal{P}_0(x)=\phi_0^2(x)=\exp(2w(x))$ (with $\int
\mathcal{P}_0(x)\,dx=1$), which is obviously non-negative. Any
positive definite initial probability density $\mathcal{P}(x,0)$
can be expanded as $\mathcal{P}(x,0)=\phi_0(x)\sum_n
c_n\phi_n(x)$, with constant coefficients $c_n$ ($n=0,1,\ldots$)
\begin{eqnarray}
c_n=\int_{-\infty}^\infty \phi_n(x)\left(\phi_0^{-1}(x)
\mathcal{P}(x,0)\right)dx.
\label{c_n}
\end{eqnarray}
Note that  $c_0= \int_{-\infty}^\infty 
\mathcal{P}(x,0)dx=1$.
At any later time $t$, the solution of the FPE is
\begin{equation}
\mathcal{P}(x,t)=\phi_0(x)\sum_n c_n \phi_n(x)\exp(-\mathcal{E}_n t).
\label{pdf}
\end{equation}

In this paper we shall be interested in the  initial profile given by the $\delta$-function
\begin{equation}
\mathcal{P}(x,t)=\delta(x-x_0).
\end{equation}
This corresponds to the situation where the particle performing the stochastic motion is initially located at the point $x_0$.
From Eqs.~(\ref{c_n}) and (\ref{pdf}) we have
\begin{equation}
c_n=\phi_0^{-1}(x_0)\phi_n(x_0),
\end{equation}
and 
\begin{equation}
\mathcal{P}(x,t)=\phi_0(x)\phi_0^{-1}(x_0)\sum_n \phi_n (x_0) \phi_n(x) \exp(-\mathcal{E}_n t).
\label{pdf-d}
\end{equation}

We now consider exactly solvable  FPEs associated with the newly discovered
exceptional orthogonal polynomials.    Only the singled-indexed exceptional orthogonal polynomials will be considered here.
There are four basic families of the single-indexed exceptional orthogonal polynomials: two associated with the Laguerre types 
(termed the L1 and L2 types) ,
 and the other two with the Jacobi types (termed the J1 and J2 types).
Quantal systems associated with the Laguerre-type exceptional orthogonal polynomials are related to the radial oscillator, 
while the Jacobi-types are related to the  Darboux-P\"oschl-Teller potential.
The corresponding FPEs describe the
generalized Rayleigh process \cite{GAR}, and the generalized 
Jacobi process, respectively.

The eigenfunctions of the corresponding Schr\"odinger equations 
are given by 
\begin{equation} 
\phi_{\ell,n}(x;\bm{\lambda})=N_{\ell,n}(\bm{\lambda})\phi_{\ell}(x;\bm{\lambda})P_{\ell,n}(\eta(x);\bm{\lambda}),
\end{equation}
where $\eta(x)$ is a function of $x$, $\bm{\lambda}$ is a set of parameters of the systems, $N_{\ell,n}(\bm{\lambda})$ is a normalization constant, 
$\phi_{\ell}(x;\bm{\lambda})$ is the asymptotic factor, and $P_{\ell,n}(x;\bm{\lambda})$ is the polynomial part of the wave function,
which is given by the exceptional orthogonal polynomials.
When $\ell=0$, $P_{0,n}(x;\bm{\lambda})$ are simply the ordinary classical orthogonal polynomials of the corresponding quantal systems, 
with $P_{0,0}(x;\bm{\lambda})={\rm ~constant}$.  When $\ell>0$, $P_{\ell,n}(x;\bm{\lambda})$ is a polynomial of degree $\ell+n$. 
The PDF  (\ref{pdf-d}) for these generalized systems are
\begin{eqnarray}
\mathcal{P}(x,t)=&&\phi_\ell^2(x;\bm{\lambda}) P_{\ell,0}(\eta;\bm{\lambda}) P^{-1}_{\ell,0}(\eta_0\bm{\lambda})\n\\
&&\times  \sum_{n=0}^\infty N_{\ell,n}(\bm{\lambda})^2 P_{\ell,n}(\eta;\bm{\lambda})P_{\ell,n}(\eta_0;\bm{\lambda})\exp(-\mathcal{E}_n(\bm{\lambda}) t).
\label{pdf-d-ell}
\end{eqnarray}

In the next two sections,  we shall consider the Laguerre and Jacobi systems separately.




\section{FPEs with Exceptional $X_\ell$ Laguerre polynomials}

\subsection{Deformed drift potential and probability density function}

We now consider the
exceptional $X_{\ell}$ Laguerre polynomials, which appear in the
deformed radial oscillator potentials.  The original radial
oscillator potential is generated by the prepotential
 \be
w_0(x;g)=-\frac{ x^2}{2}+g\log x,~~0<x<\infty.
 \label{w0}
 \ee
The exactly solvable Schr\"odinger equations associated with the deformed radial oscillator potential 
are defined by the prepotentials ($\ell=1,2,\ldots$)
\begin{equation}
w_\ell(x;g)= -\frac{x^2}{2}+(g+\ell)\log x
+\log\frac{\xi_\ell(\eta;g+1)}{\xi_\ell(\eta;g)}.\label{wl-osc}
\end{equation}
Here  $\bm{\lambda}=g>0$ and 
$\eta(x)\equiv  x^2$ and $\xi_\ell(\eta;g)$ is a deforming function.
 It turns out
there are two possible sets of deforming functions
$\xi_\ell(\eta;g)$, thus giving rise to two sets of infinitely
many exceptional Laguerre polynomials, termed L1 and L2 type
\cite{OS1,HOS}. These $\xi_\ell$ are given by
 \be
  \xi_{\ell}(\eta;g)=
  \left\{
  \begin{array}{ll}
  L_{\ell}^{(g+\ell-\frac32)}(-\eta)&:\text{L1}\\
  L_{\ell}^{(-g-\ell-\frac12)}(\eta)&:\text{L2}
  \end{array}\right. .
  \label{xiL}
\ee

For both L1 and L2 type Laguerre systems, the eigen-energies are
$\mathcal{E}_{\ell,n}(g)=\mathcal{E}_n(g+\ell)=4n$, which
are independent of $g$ and $\ell$. Hence the deformed radial
oscillator is iso-spectral to the ordinary radial oscillator. The
eigenfunctions are given by
 \be
\phi_{\ell,n}(x;g)=N_{\ell,n}(g) \phi_\ell(x;g) P_{\ell,n}(\eta;g),~~~
\phi_\ell(x;g)&\equiv & \frac{e^{-\frac{1}{2}
x^2}x^{g+\ell}}{\xi_{\ell}(x^2;g)}, \label{phi-l}
\ee
where the corresponding exceptional Laguerre polynomials
$P_{\ell,n}(\eta;g)$ ($\ell=1,2,\ldots$, $n=0,1,2,\ldots$) can be
expressed as a bilinear form of the classical associated Laguerre polynomials
$L^\alpha_n(\eta)$ 
and the deforming polynomial $\xi_\ell(\eta;g)$, as given in \cite{HOS}:
\be
P_{\ell,n}(\eta;g)=
  \left\{
  \begin{array}{ll}
  \xi_{\ell}(\eta;g+1)P_n(\eta;g+\ell-1)
  -\xi_{\ell}(\eta;g)\partial_{\eta}
  P_n(\eta;g+\ell-1)&:\text{L1}\\[2pt]
  (n+g+\frac12)^{-1}\bigl\{(g+\frac12)
  \xi_{\ell}(\eta;g+1)P_n(\eta;g+\ell+1)\\
  \phantom{(n+g+\frac12)^{-1}\bigl(}\ \quad
  +\eta\xi_{\ell}(\eta;g)\partial_{\eta}P_n(\eta;g+\ell+1)
  \bigr\}&:\text{L2}.
 \end{array}\right. ,
  \label{XL}
\ee
where
$P_n(\eta;g)\equiv
  L_n^{(g-\frac{1}{2})}(\eta)$.
 The $X_{\ell}$ polynomials
$P_{\ell,n}(\eta;g)$ are degree $\ell+n$ polynomials in $\eta$ and
start at degree $\ell$: $P_{\ell,0}(\eta;g)=\xi_{\ell}(\eta;g+1)$.
They are orthogonal with respect to certain weight functions,
which are deformations of the weight function for the Laguerre
polynomials (for details, see \cite{HOS}).
From the orthogonality relation of $P_{\ell,n}(\eta;g)$, we get the normalization constants \cite{HOS}
\begin{equation}
  N_{\ell,n}(g)=
  \left\{
  \begin{array}{ll}
  \left[\frac{2n! (n+g+\ell-\frac12)}{(n+g+2\ell-\frac12)\Gamma (n+g+\ell+\frac12)}\right]^{\frac12}&:\text{L1}\\[2pt]
 \left[\frac{2n! (n+g+\frac12)}{(n+g+\ell+\frac12)\Gamma (n+g+\ell+\frac12)}\right]^{\frac12}&:\text{L2}
  \end{array}
  \right. .
\end{equation}

The drift coefficient and PDF of the FPE  generated by the prepotential $w_\ell(x)$ in eq.(\ref{wl-osc}) are
\be
D^{(1)}(x)=-2x+ 2\frac{g+\ell}{x}+4x\left(\frac{\partial_\eta\xi_\ell(\eta;g+1)}{\xi_\ell(\eta;g+1)}-\frac{\partial_\eta\xi_\ell(\eta;g)}{\xi_\ell(\eta;g)}\right)
\label{D1-L}
\ee
and
\begin{eqnarray}
\mathcal{P}(x,t)&=&\phi_\ell^2(x;g) P_{\ell,0}(x^2;g) P^{-1}_{\ell,0}(x_0^2;g)\nonumber\\
&&\times  \sum_{n=0}^\infty N_{\ell,n}^2 P_{\ell,n}(x^2;g)P_{\ell,n}(x_0^2;g)\exp(-\lambda_n t)
\label{pdf-L}
\end{eqnarray}
In the limit $\ell\to 0$ for $g=1/2$,  we have $\xi_\ell(x;g+1)\to 0$ and $\xi_\ell(x;g)\to 0$, and eqs.~(\ref{D1-L}) and (\ref{pdf-L}) become
\be
D^{(1)}&=&-2x+\frac{1}{x},\\
\mathcal{P}(x,t)&=&2\, x\, e^{-x^2}\sum_{n=0}^\infty L_n(x^2) L_n(x_0^2)e^{-4nt},
\ee
where $L_n(x)$ are the ordinary  Laguerre polynomials. 
These are just the drift coefficient (with the diffusion coefficient being one) and the PDF for the Rayleigh 
process as given in \cite{GAR}.

\subsection{Numerical results}

Below we shall present some numerical results in order to see how the deforming function $\xi_\ell(\eta; g)$ modifies the behavior
of the drift-diffusion system. 
Since the diffusion coefficient here is fixed, and the deformation of the system is mainly through the change in the
drift coefficient, it suffices to see how the drift potential is deformed as $\ell$ changes. This we plot in Fig.~1 for the L1 type
Laguerre systems.  The L2 systems behave similarly, and will not be depicted here. 
Four sets of values of $g$ are shown, namely, $g=0.1, 0.5, 1.0$ and $5.0$.  For each value of $g$, we plot $U_\ell(x)$
with $\ell=0$ (the undeformed case), $1$ and $5$.  

Since the drift coefficient is related to the negative slope of $U_\ell(x)$,
one can gain some qualitative understanding of how the peak of the probability density moves just from the sign of the
slope of $U_\ell(x)$.  The reason is as follows. Let us consider the current density $J(x,t)$ near the peak, where the slope
of the probability density function is very small, $\partial P(x,t)/\partial x\approx 0$.  From eq.~(\ref{J}), one has
\begin{equation}
J(x,t)_{\rm peak}\approx -\left(\frac{\partial U_\ell(x)}{\partial x}\, P(x,t)\right)_{\rm peak}.\label{J-peak}
\end{equation}
As $P(x,t)>0$ near the peak, eq.~(\ref{J-peak}) implies that the peak tends to move to the right (left) when it is at a position $x$ such that $\partial U_\ell(x)/\partial x$ is
negative (positive), until the stationary distribution is reached.

From Fig.~1 one sees that for the values of $g$ considered, the presence of parameter $\ell$ generally makes the slope of $U_\ell (x)$ more
negative near the left wall $x=0$. So in  the deformed systems, the peaks tend to move faster to the right.  Also, there can be
sign change of the slope of the drift potential  in certain region
near the left wall. In such region,  the peak of the probability density function will move in different directions for different  $\ell$.
To illustrate this, we consider the system with $g=0.5$.  As mentioned before, the $\ell=0$ case with $g=0.5$  correspond to the well-known Rayleigh process.
From Fig.~1 we see that at $x_0=1.2$, the sign of the slope of $U_5(x)$ is different from those of  $U_0(x)$ and $U_1(x)$. Thus one expects that
for the initial profile $P(x,0)=\delta(x-x_0)$ with the peak initially located at $x_0=1.2$, the peak will move to the right for $\ell=5$ system, while for the other 
two values of $\ell$, the peak will move to the left.  This is indeed the case  as shown in Fig.~2.  We note here that the graphs shown in Fig.~2 for $t=2.0$ 
are already the stationary distributions of the respective cases.




\section{FPEs with Exceptional Jacobi polynomials}

We now present new exactly solvable FPEs related to the exceptional Jacobi polynomials.  We shall 
call these the generalized Jacobi processes. 
Here we will be brief,
as the method of construction of these systems is the same as that discussed in the previous section.

With appropriate redefinition, one can always set the domain of the basic variable to be $x\in [0,\pi/2]$.
The trigonometric  Darboux-P\"oschl-Teller potential has two parameters $\bm{\lambda}=(g,h)$ ($g>0$, $h>0$) and is defined by the 
prepotential ($\ell=1,2,\ldots$)
\begin{align}
w_0(x;g,h)= g\log\sin x+h\log\cos x,\quad 0<x<\frac{\pi}{2}.
\end{align}
The corresponding deformation of the system is given by the prepotential
\be
w_\ell(x;g,h)= (g+\ell)\log\sin x +(h+\ell)\log\cos x
+\log\frac{\xi_\ell(\eta;g+1,h+1)}{\xi_\ell(\eta;g,h)},
\ee
where $\eta(x)=\cos 2x$. The deforming function is defined by
\be
 \xi_{\ell}(\eta;g,h)=
  \left\{
  \begin{array}{ll}
  P_{\ell}^{(g+\ell-\frac32,-h-\ell-\frac12)}(\eta),\ \ g>h>0&:\text{J1}\\
  P_{\ell}^{(-g-\ell-\frac12,h+\ell-\frac32)}(\eta),\ \ h>g>0&:\text{J2}
  \end{array}\right. ,\n
\ee
where $P_n^{(\alpha,\beta)}(x)$ are
the classical Jacobi polynomials.

The drift coefficients of these systems are
\be
D^{(1)}(x)=2(g+\ell)\cot x - 2(h+\ell)\tan x-2\sin( 2x) \left(\frac{\partial_\eta\xi_\ell(\eta; g+1,h+1)}{\xi_\ell(\eta; g+1,h+1)}
-\frac{\partial_\eta\xi_\ell(\eta; g,h)}{\xi_\ell(\eta; g,h)}\right).
\label{D1-L}
\ee
The PDF of these systems are again given by eq.~(\ref{pdf-d-ell}).   The basic data needed in the formula are
 \cite{HOS}:
\begin{align} 
&\mathcal{E}_{\ell,n}(g,h)=4n(n+g+h+2\ell),\n\\
&\phi_{\ell,n}(x;g,h)=\phi_{\ell}(x;g,h)P_{\ell,n}(\eta;g,h),\quad 
\phi_{\ell}(x;g,h)=\frac{(\sin x)^{g+\ell}(\cos x)^{h+\ell}}{ \xi_{\ell}(\eta;g,h)},\n\\
 & N_n(g,h)=\left[\frac {2\,n!\,(2n+g+h)\Gamma(n+g+h)}{\Gamma(n+g+\frac12)\Gamma(n+h+\frac12)}\right]^\frac12,\n\\
& N^2_{\ell,n}(g,h)=
  \,N^2_n(g+\ell,h+\ell)\times
  \left\{
  \begin{array}{ll}
  \frac{(n+h+\frac12)(n+g+\ell-\frac12)}{(n+h+\ell+\frac12)(n+g+2\ell-\frac12)}
  &:\text{J1}\\[4pt]
  \frac {(n+g+\frac12)(n+h+\ell-\frac12)}{(n+g+\ell+\frac12)(n+h+2\ell-\frac12)}
 &:\text{J2}
  \end{array}\right.,\\
 &   P_{\ell,n}(\eta;g,h)=
  \left\{
  \begin{array}{ll}
  (n+h+\frac12)^{-1}\bigl\{
  (h+\frac12)\xi_{\ell}(\eta;g+1,h+1)
  P_n^{(g+\ell-\frac32,h+\ell+\frac12)}(\eta)&\\
  \phantom{(n+h+\frac12)^{-1}\bigl(}
  +(1+\eta)\xi_{\ell}(\eta;g,h)
  \partial_{\eta}P_{n}^{(g+\ell-\frac32,h+\ell+\frac12)}(\eta)\bigr\}
  &:\text{J1}\\[2pt]
  (n+g+\frac12)^{-1}\bigl\{
  (g+\frac12)\xi_{\ell}(\eta;g+1,h+1)
  P_n^{(g+\ell+\frac12,h+\ell-\frac32)}(\eta)&\\
  \phantom{(n+g+\frac12)^{-1}\bigl(}
  -(1-\eta)\xi_{\ell}(\eta;g,h)
  \partial_{\eta}P_{n}^{(g+\ell+\frac12,h+\ell-\frac32)}(\eta)\bigr\}
  &:\text{J2}
  \end{array}\right.. \hspace{2mm} \n
\end{align}

As explained in \cite{OS1,HOS}, the exceptional J1 and J2 orthogonal polynomials are 
`mirror images' of each other, reflecting the parity property 
$P_n^{(\alpha,\beta)}(-\eta)=(-1)^nP_n^{(\beta,\alpha)}(\eta)$ of 
the Jacobi polynomial.


Below we discuss some numerical results for the J2 systems.  In Fig.~3 we plot the drift potentials
$U_\ell(x)$ for various sets of parameters.

Figs.~3(a) and 3(b)  show how $U_\ell(x)$ changes as $\ell$ changes at
fixed values of $g$ and $h$.  In (a) the 
difference between $h$ and $g$ is small,  and the graphs of $U_\ell(x)$ appear nearly symmetric with
respect to the mid-point $x=\pi/4$, but with the absolute 
value of the slope becoming 
larger at higher $\ell$. So the probability
density is pushed faster towards the central region of the well at higher $\ell$.  
In (b) $h$ is much larger than $g$, and one sees that there can be
difference in the signs of the slopes of the drift potential near the left wall.  The effect of the sign change is depicted in Fig.~4.

In Figs.~3(c) and 3(d), we show the change in $U_\ell(x)$ (at fixed value of $\ell$)
when either $g$ or $h$ is changed while the other is kept constant.   It is seen that 
$U_\ell(x)$ becomes steeper near the left (right) wall as $g$ ($h$) becomes larger at fixed $h$ ($g$). 
This implies  the  probability density
function near the left  (right) wall is pushed faster towards the right (left)  at larger $g$ ($h$) at fixed $h$ ($g$). 

In Fig.~4 the time evolution of the probably density function is depicted for the three cases shown in Fig.~3(b).
It has been noted that at some points , the slope of $U_\ell(x)$ can have different signs at different values of $\ell$ (at fixed $g$ and $h$).  For instance, 
at  $x_0=0.3$, the slope is positive for $\ell=0$, and are negative for at $\ell=10$ and $15$. If the initial profile peaks at $x_0=0.3$,
one expects from the signs of the slopes of $U_\ell(x)$ that  the peak will move towards the  left wall for the non-deformed system, and 0towards the right
for the other two deformed cases.  Fig.~4 shows that this is indeed the case.




\section{Summary}

We have presented four types of infinitely
many exactly solvable FPE's which are the generalised, or deformed 
versions of the Rayleigh process and the Jacobi process.   They are related to
the newly discovered exceptional orthogonal polynomials.  These new 
polynomials have the remarkable properties that they  start with degree $\ell=1,2\ldots,$
polynomials instead of a constant, and yet they still form complete sets with respect to some positive-definite measure.
In some sense they are  deformation of the classical Laguerre and Jacobi polynomials. 
The quantal systems that involve these new polynomials are the deformed radial oscillator and the
 trigonometric  Darboux-P\"oschl-Teller potential. 
 Using the transformation between FPE and the Schr\"odinger equation, we obtained the corresponding FPE's,
which are deformations of the Rayleigh process and the Jacobi process.
We have shown numerically how the deformation modify the drift coefficient of the FPE, and hence the evolution of the probability density function.
It gives the possibility to manipulate evolution of stochastic processes by suitable choice of a deforming function. 


\begin{acknowledgments}

This work is supported in part by the
National Science Council (NSC) of the Republic of China under
Grants NSC-99-2112-M-032-002-MY3 and NSC-101-2918-I-032-001. 
The final part of the work was done 
during CLH's visit to  the Yukawa Institute for
Theoretical Physics (YITP), Kyoto University, and he would like to
thank the staff and members of YITP for the hospitality extended
to him, and to R. Sasaki for helpful discussion on exceptional orthogonal
polynomials.  

\end{acknowledgments}

\newpage

\newpage



\begin{figure}[ht] \centering
\includegraphics*[width=7.5cm,height=7.5cm]{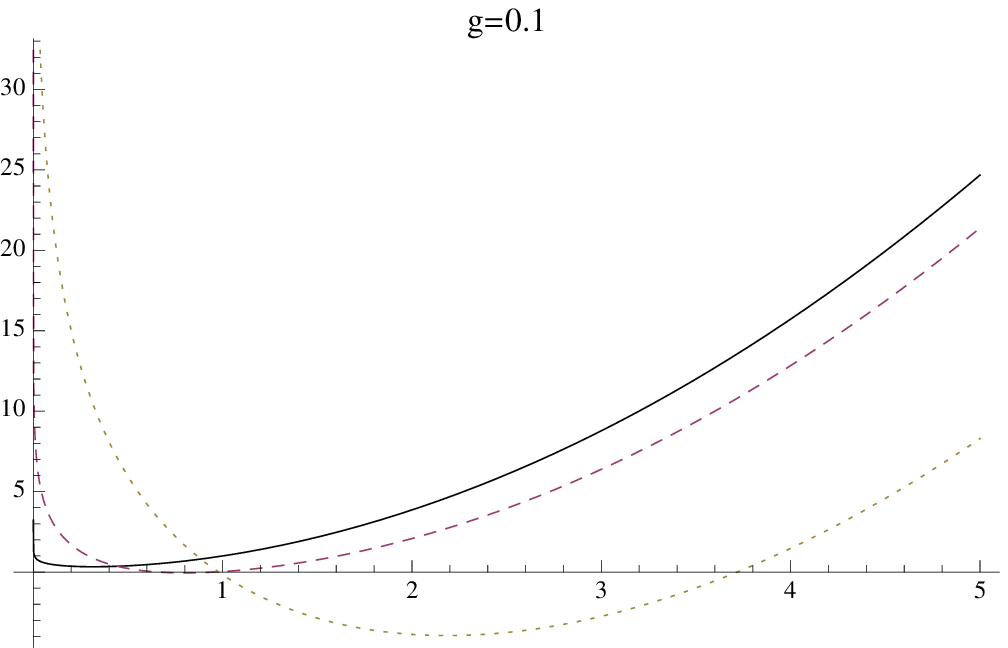}\hspace{1cm}
\includegraphics*[width=7.5cm,height=7.5cm]{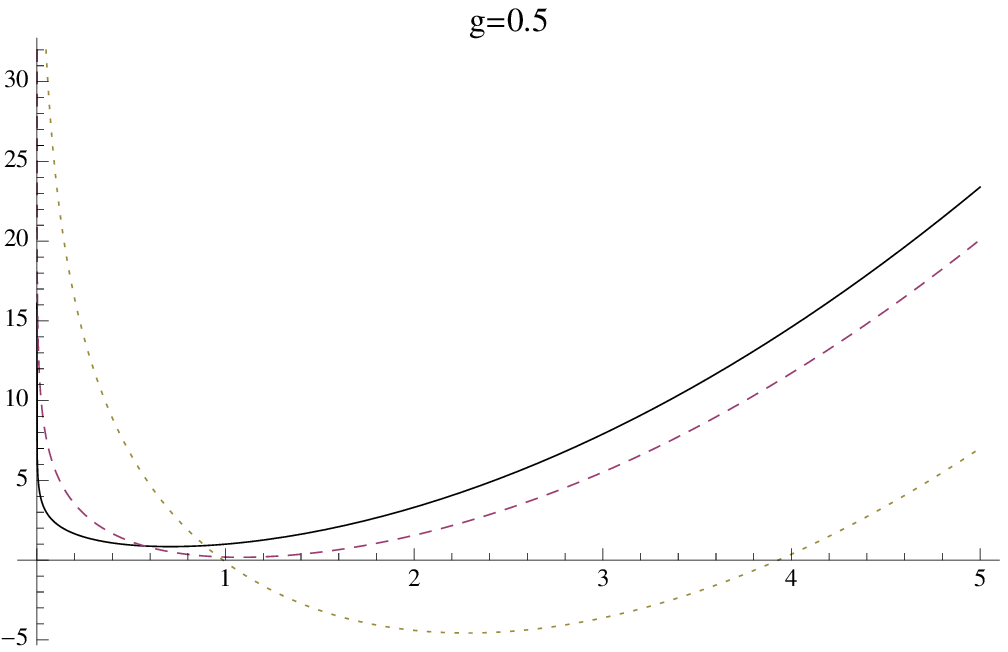}\vspace{3cm}
\includegraphics*[width=7.5cm,height=7.5cm]{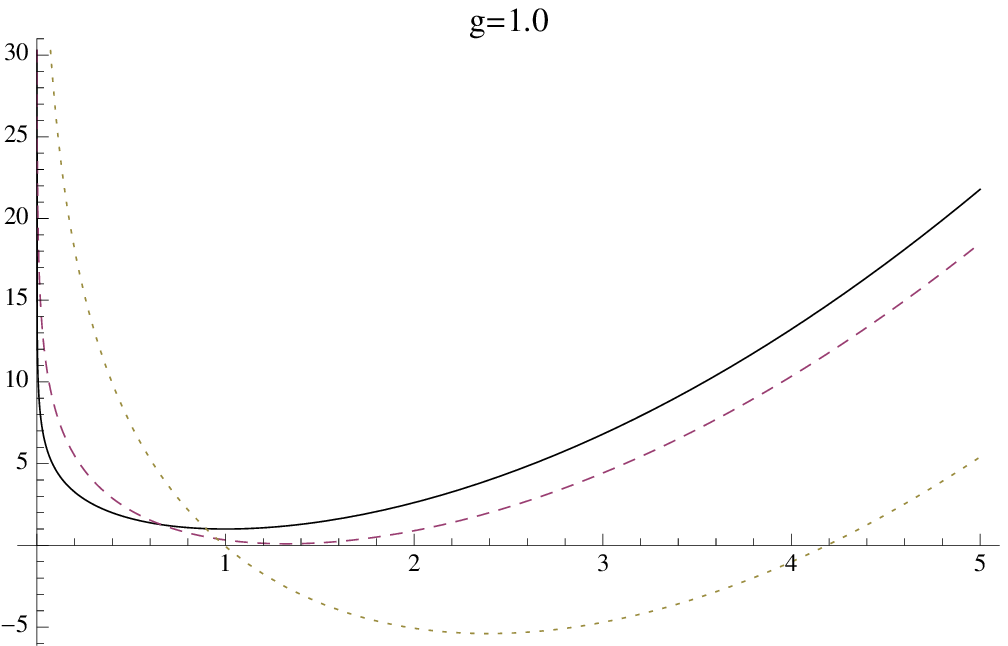}\hspace{1cm}
\includegraphics*[width=7.5cm,height=7.5cm]{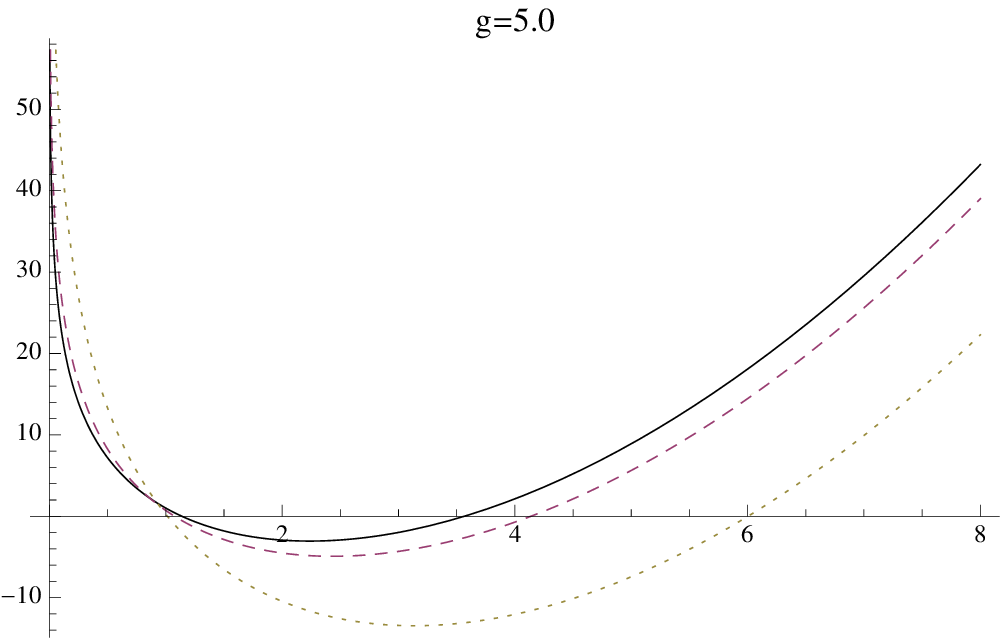}
\caption{ Plots of $U_\ell (x)$ versus $x$ for the L1 type FPE with different values of $g$ and $\ell=0$ (solid curve), $1$ (dashed curve) and $5$ (dotted curve)} \label{Fig1}
\end{figure}

\begin{figure}[ht] \centering
\includegraphics*[width=7.5cm,height=7.5cm]{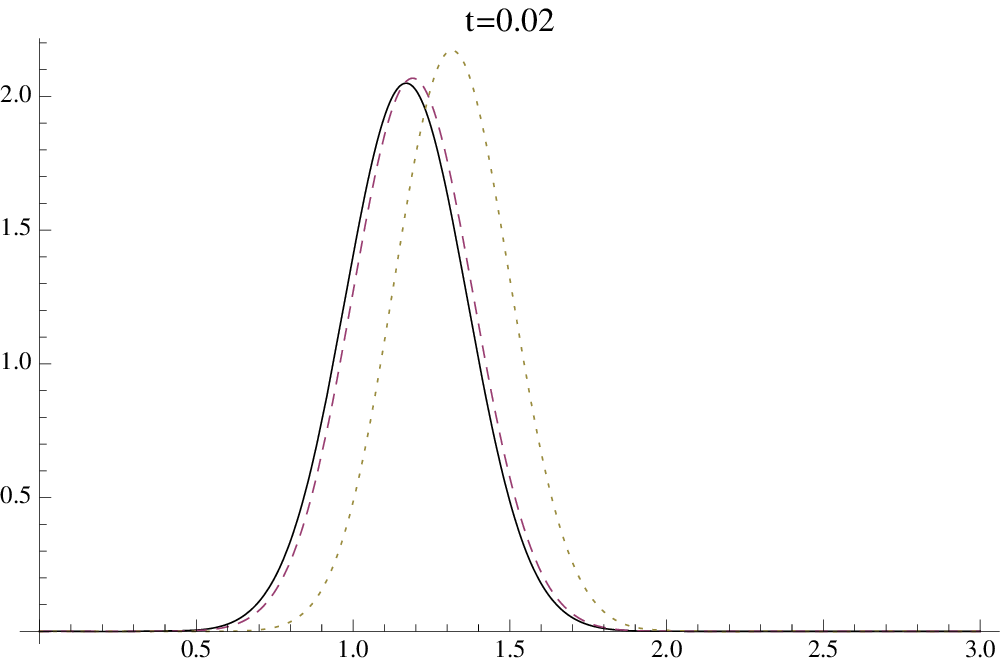}\hspace{1cm}
\includegraphics*[width=7.5cm,height=7.5cm]{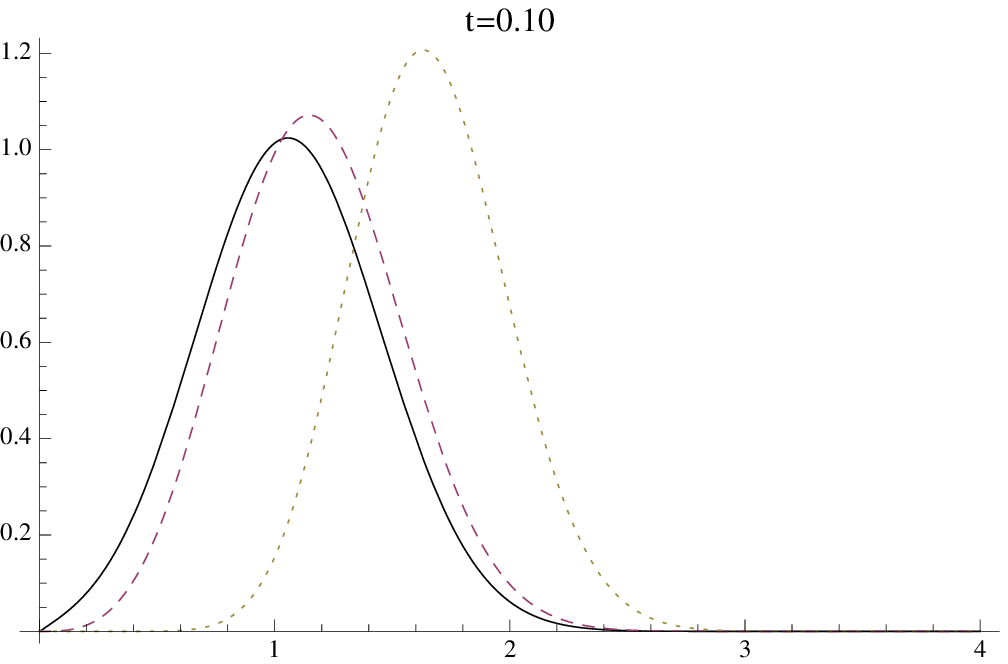}\vspace{3cm}
\includegraphics*[width=7.5cm,height=7.5cm]{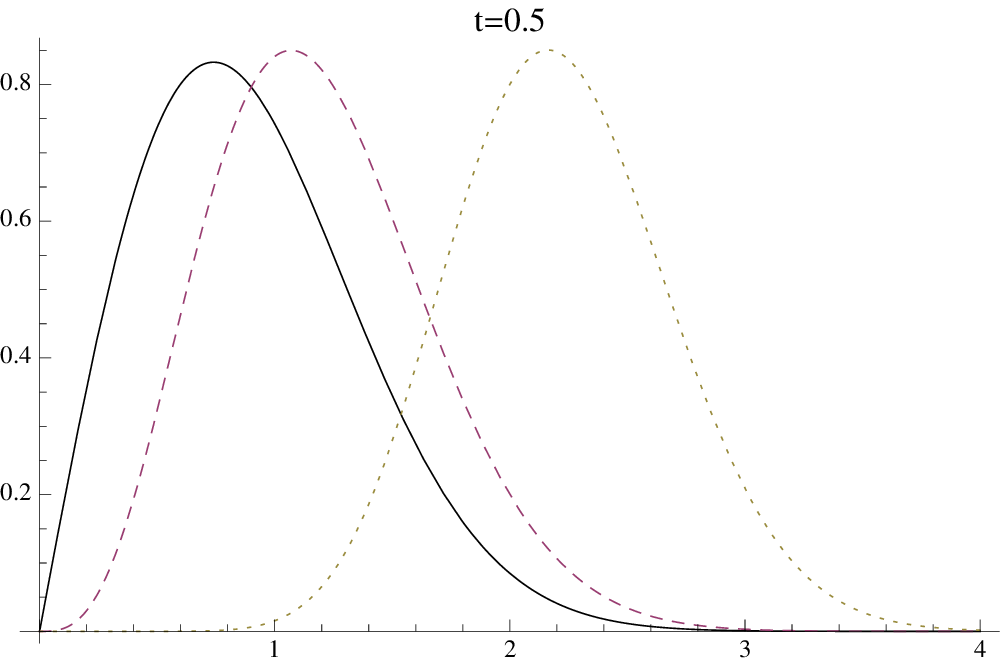}\hspace{1cm}
\includegraphics*[width=7.5cm,height=7.5cm]{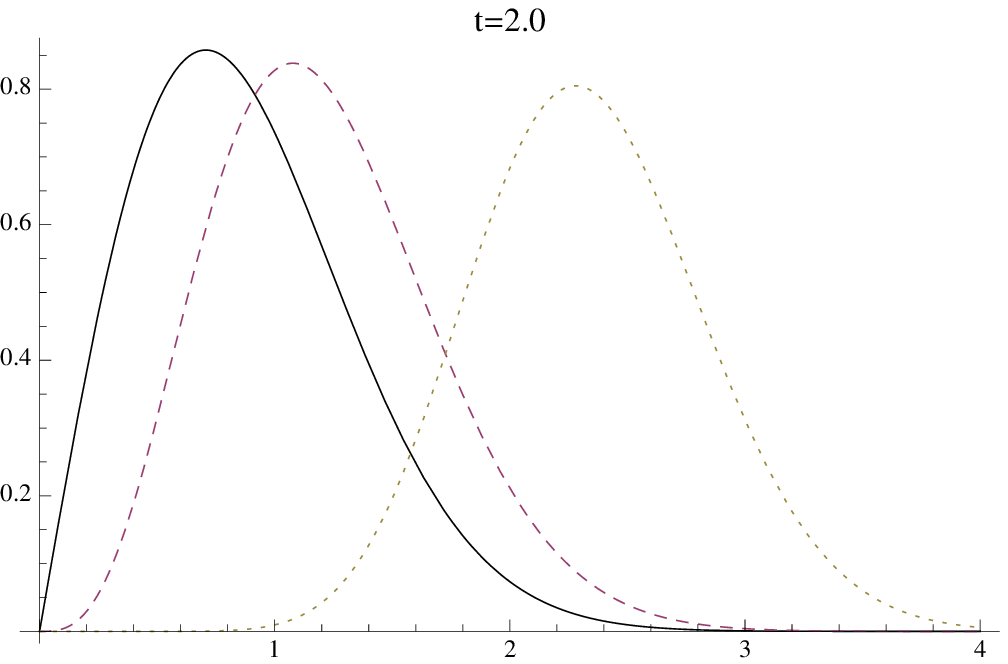}
\caption{Plots of  $\mathcal{P}(x,t)$ versus $x$ for the L1 FPE with different values of $t$ and $g=0.5, x_0=1.2$, 
$\ell=0$ (solid curve), $1$ (dashed curve) and $5$ (dotted curve).  Eighty terms in the series in eq.~(\ref{pdf-L}) were  used.
The curves shown in the last sub-figure are the stationary distributions for the respective value of $\ell$.
} \label{Fig2}
\end{figure}

\begin{figure}[ht] \centering
\includegraphics*[width=7.5cm,height=7.5cm]{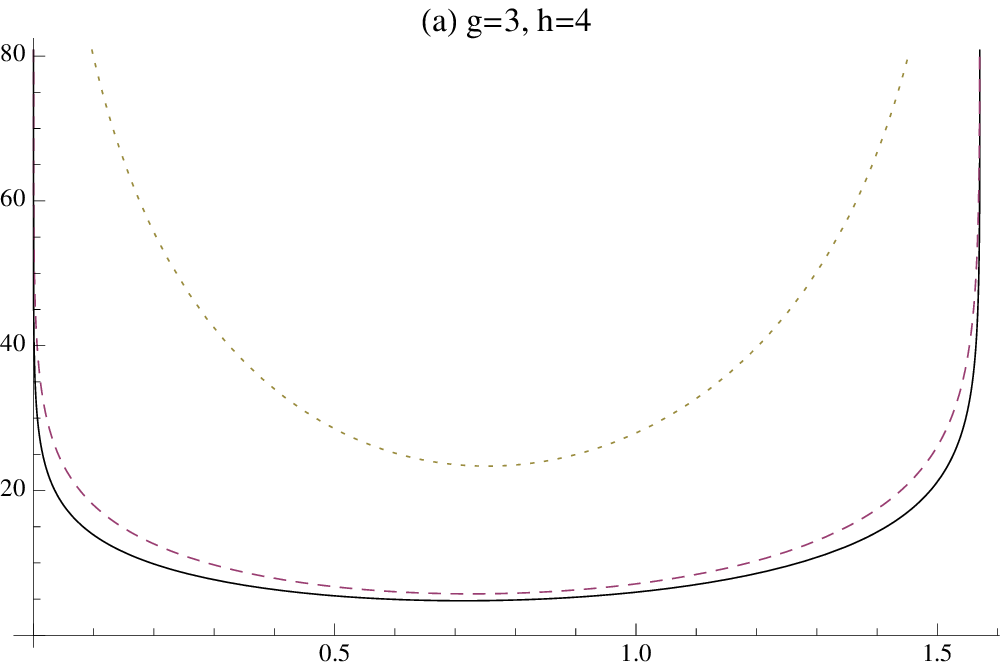}\hspace{1cm}
\includegraphics*[width=7.5cm,height=7.5cm]{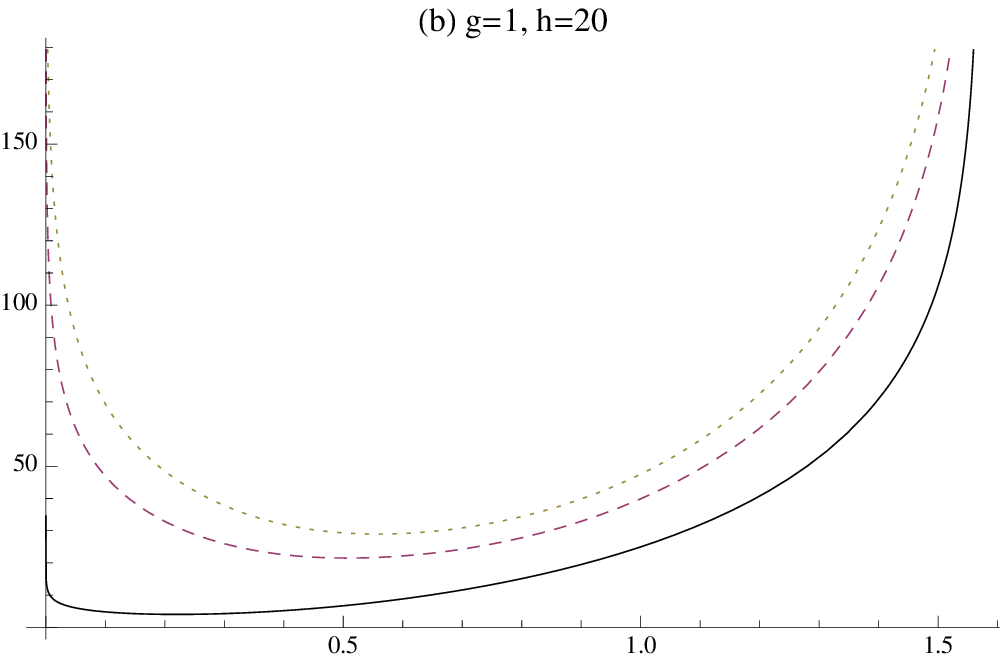}\vspace{3cm}
\includegraphics*[width=7.5cm,height=7.5cm]{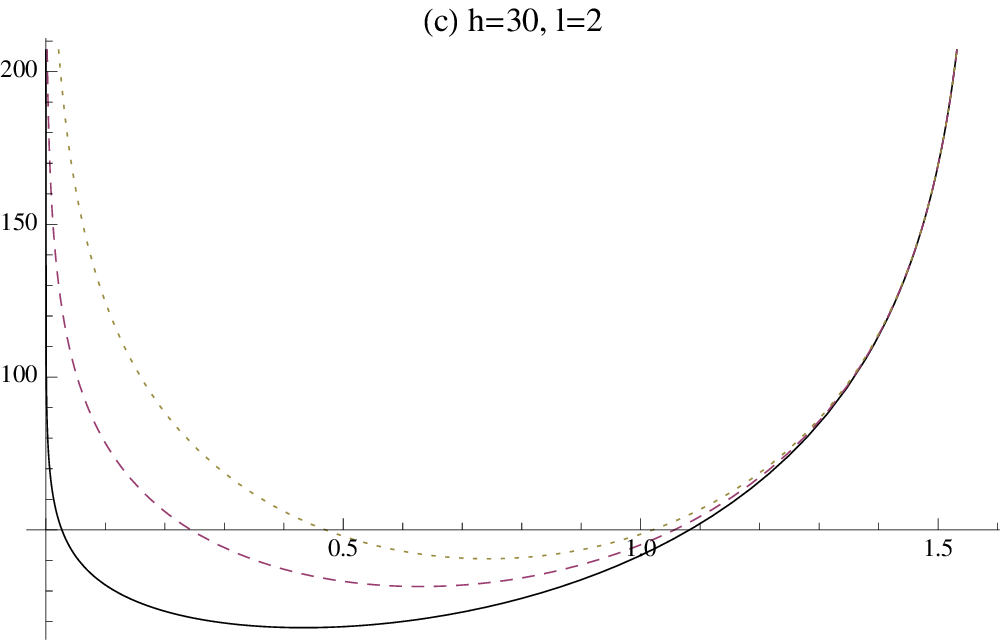}\hspace{1cm}
\includegraphics*[width=7.5cm,height=7.5cm]{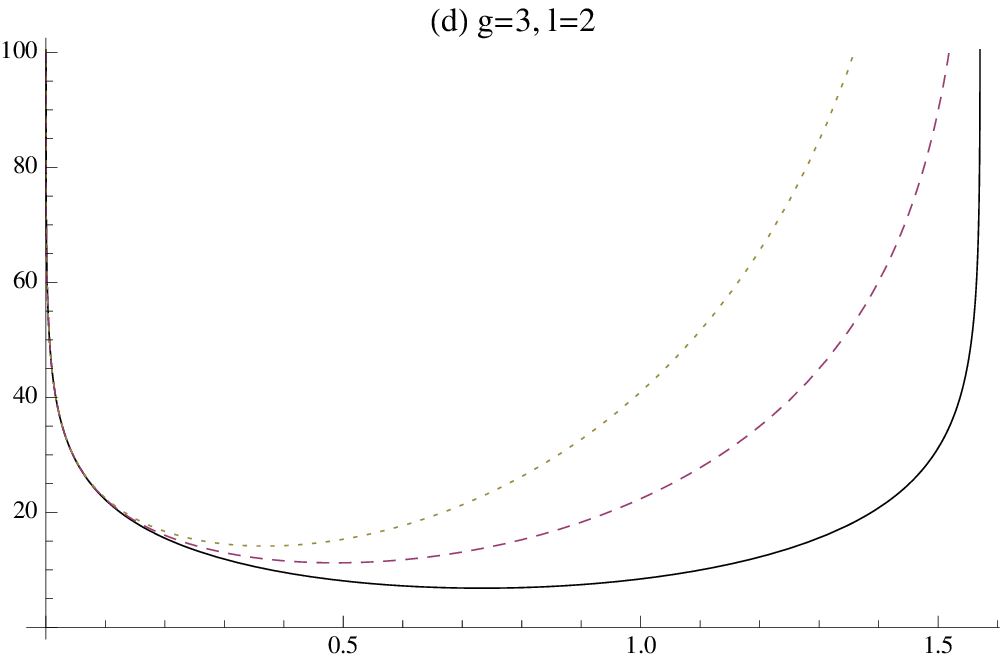}
\caption{Plots of  $U_\ell (x)$ versus $x$ for J2 type FPE for : (a) $g=3, h=4$,  $\ell=0$ (solid curve), $1$ (dashed curve) and $15$ (dotted curve); 
(b)  $g=1, h=20$,  $\ell=0$ (solid curve), $10$ (dashed curve) and $15$ (dotted curve); 
(c)  $h=30, \ell=2$,  $g=5$ (solid curve), $15$ (dashed curve) and $25$ (dotted curve);  
and 
(d)  $g=3, \ell=2$,  $h=4$ (solid curve), $15$ (dashed curve) and $30$ (dotted curve)}.\label{Fig3}
\end{figure}

\begin{figure}[ht] \centering
\includegraphics*[width=7.5cm,height=7.5cm]{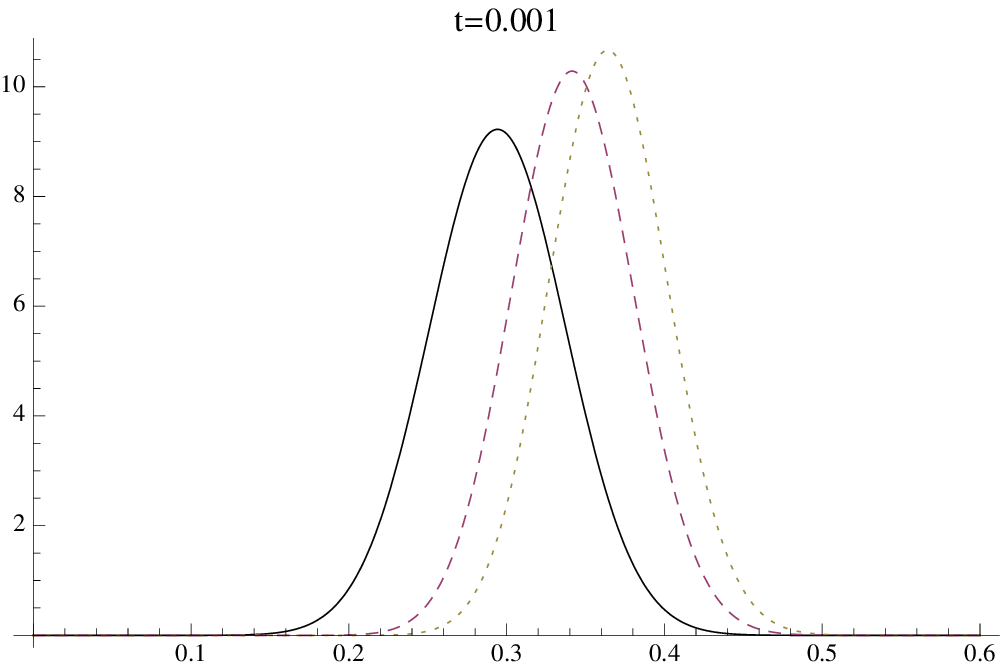}\hspace{1cm}
\includegraphics*[width=7.5cm,height=7.5cm]{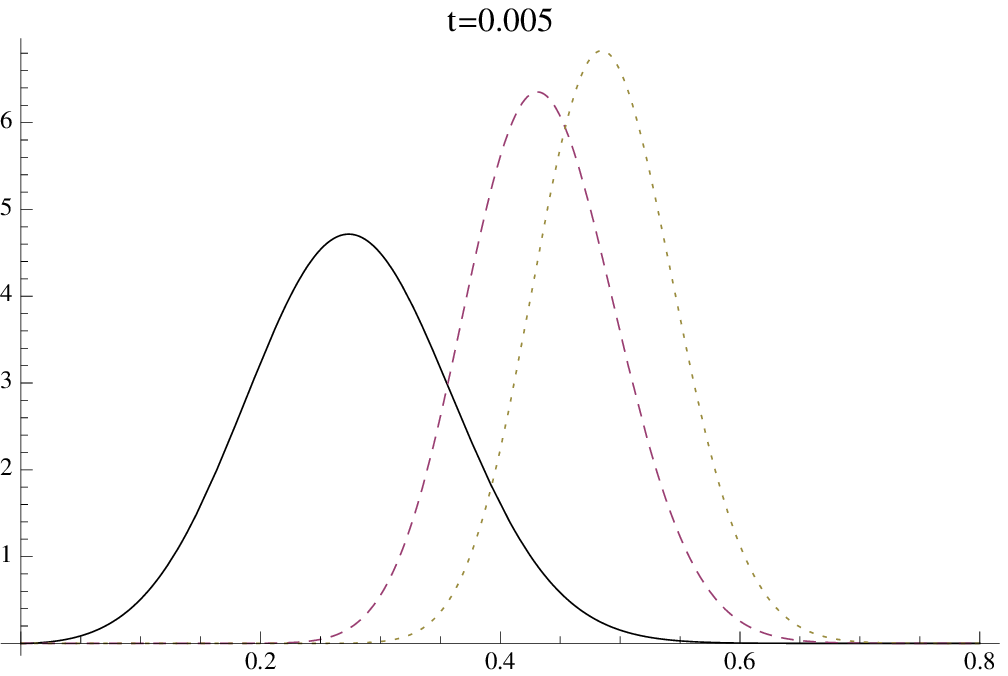}\vspace{3cm}
\includegraphics*[width=7.5cm,height=7.5cm]{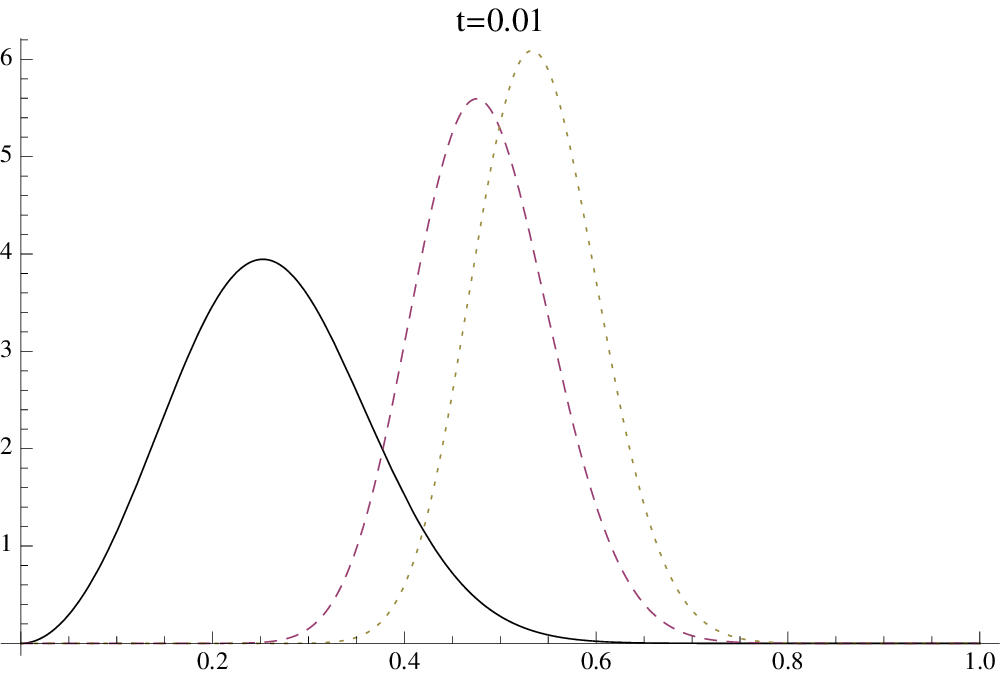}\hspace{1cm}
\includegraphics*[width=7.5cm,height=7.5cm]{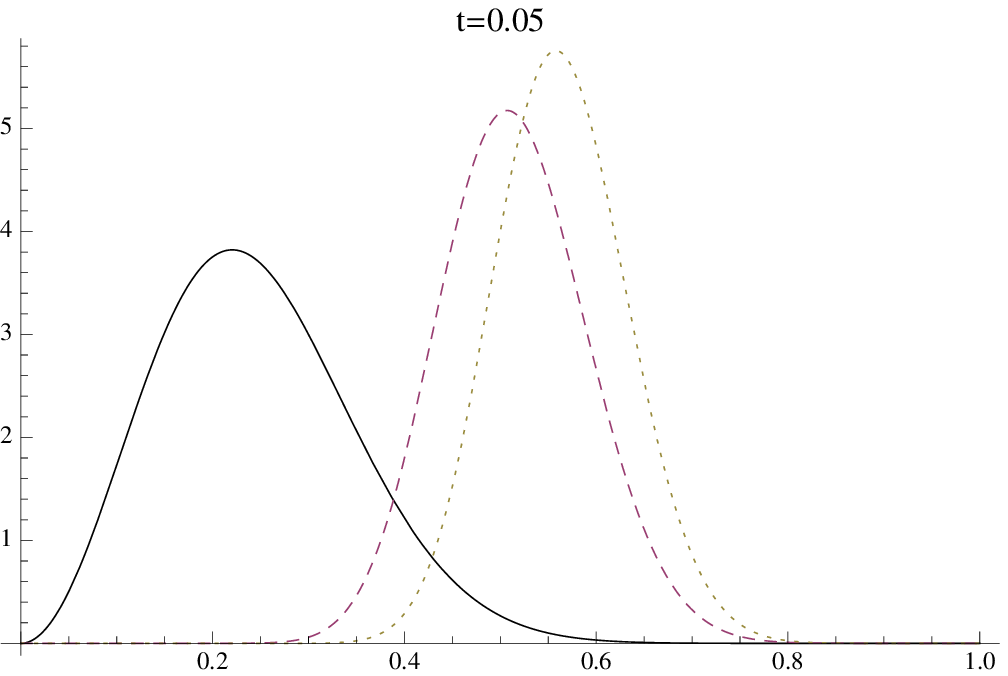}
\caption{Plots of  $\mathcal{P}(x,t)$ versus $x$ for the J2 FPE with different values of $t$ and $g=1,  h=20, x_0=0.3$, 
$\ell=0$ (solid curve), $10$ (dashed curve) and $15$ (dotted curve).  Fifty terms in the series in eq.~(\ref{pdf-d-ell}) were  used.
The curves shown in the last sub-figure are the stationary distributions for the respective value of $\ell$.} \label{Fig4}
\end{figure}

\end{document}